\documentclass[aps,prl,twocolumn,superscriptaddress,showpacs]{revtex4}

\usepackage{graphics}
\usepackage{bm}

\usepackage{amsmath}    
\usepackage{amssymb}    
\usepackage{graphicx}   
\def\registered{{\ooalign{\hfil\raise .00ex\hbox{\scriptsize R}\hfil\crcr\mathhexbox20D}}}

\begin{document}

\title {A quantitative investigation of the effect of a close-fitting superconducting shield on the coil-factor of a solenoid}
\thanks{To be submitted to Applied Physics Letter}

\author{M. Aaroe}
\affiliation{DTU Physics, B309, Technical University of
Denmark, DK-2800 Lyngby, Denmark}\email{aaroe@fysik.dtu.dk}
\author{R. Monaco}
\affiliation{Istituto di Cibernetica del CNR, 80078, Pozzuoli, Italy
and Unit$\grave{\rm a}$ INFM $­$ Dipartimento di Fisica, Universit$\grave{\rm a}$ di Salerno, 84081 Baronissi, Italy}\email{roberto@sa.infn.it}\author{V.P. Koshelets}
\affiliation{Institute of Radio Engineering and Electronics, Russian
Academy of Science, Mokhovaya 11, B7, 125009, Moscow, Russia}
\author{J. Mygind}
\affiliation{DTU Physics, B309, Technical University of
Denmark, DK-2800 Lyngby, Denmark}

\date{\today}

\begin{abstract}

Superconducting shields are commonly used to suppress external magnetic interference. We show, that an error of almost an order of magnitude can occur in the coil-factor in realistic configurations of the solenoid and the shield. The reason is that the coil-factor is determined by not only the geometry of the solenoid, but also the nearby magnetic environment. This has important consequences for many cryogenic experiments involving magnetic fields such as the determination of the parameters of Josephson junctions, as well as other superconducting devices. It is proposed to solve the problem by inserting a thin sheet of high-permeability material, and the result numerically tested.
\end{abstract}
\maketitle

\newpage
Many experiments characterizing superconductors and superconducting
devices involve applying a magnetic field. One typical class of such
experiments is the characterization of Josephson
junctions\cite{Broom}.

Superconducting shields are unsurpassed to prevent extraneous AC and
DC magnetic fields, e.g., high frequency magnetic noise and the
Earth's magnetic field, from affecting magnetically delicate cryogenic
instruments and experiments. However, in order to measure the
magnetic properties of specimens in such setups, one has to mount
one or more solenoids inside the shield. Often the trade off between
demands for homogeneous fields and limited space places the coil in
close vicinity of the shield. It is not surprising\cite{Pourrahimi}
- but often forgotten - that a shield  which is close-fitting around
the coil may strongly deform the magnetic field lines and thus
change the coil factor, $C$.


It is common practice to use a Hall probe at room temperature to calibrate coils for magnetic measurements, even when the coils are to be used in a cryogenic environment\cite{Rowell}. From this calibration it is possible to determine the coil factor, $C$, relating the DC coil current, $I_{coil}$, to the B-field, $B_i$ in the center of the solenoid

\[
B_i = C I_{coil} \ .
\]

\noindent The problem arises when the coil is consequentially
enclosed in a superconducting magnetic shield. For an ideal
high-permeability shield, with $\mu_r \rightarrow \infty$, the
problem does not arise, as the effect of the magnetically soft
shield is to create a virtual, free space for the field lines.

\begin{figure}[ht]
        \centering
                \includegraphics[width=6.5cm]{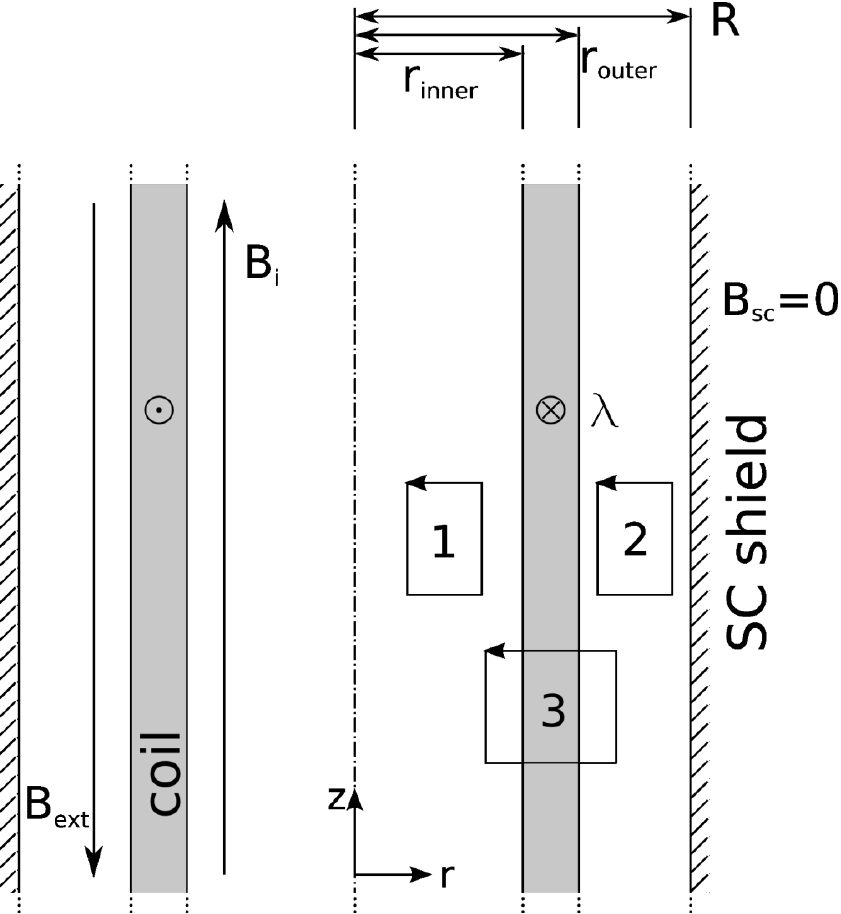}
        \caption{Illustration for the calculation of magnetic fields for infinite solenoid in an infinite cylindrical superconducting shield.}
        \label{fig:AmpereLoops}
\end{figure}

First we consider the case of the infinite solenoid in free space
and then compare to an infinite solenoid in an infinite, cylindrical
superconducting shield. In free space, the internal field of the
solenoid can be determined directly from simple, proven theoretical
expressions. If Ampere's Law is integrated along loop number 2 in
Fig. \ref{fig:AmpereLoops}, it can be seen that the field outside
the infinite solenoid, $B_{ext}$, in free space is everywhere zero,
as it must be zero at $r \rightarrow \infty$. Similarly, for loop
number 1, the field must be constant everywhere inside the volume
enclosed by the solenoid.

If we now apply Ampere's Law to loop 3 in the figure, it can be seen
that the field inside the solenoid, $B_i$, is given by

\begin{equation}
\label{eq:FieldDifference}
B_i = -B_{ext} + \mu_0\lambda \ ,
\end{equation}

\noindent where $\lambda$ is the current density (per unit length)
of the coil. Eq.(\ref{eq:FieldDifference}) is valid regardless of
the presence of a superconducting shield on the outside.

\noindent For the infinite solenoid inside a superconducting shield
we simply note, that $B_{ext} > 0$ for a close-fitting
superconducting shield ($\Delta r = R - r_{coil} \ll r_{coil}$), .
This is because the field lines must close on themselves, and now
have a limited volume in which to do so. Enclosing an infinite
solenoid in a close-fitting ideal superconducting shield ($B_{SC} =
0$) increases $B_{ext}$, while the field inside the solenoid is
reduced. Flux conservation gives:

\begin{equation}
\label{eq:FluxConservation}
A_{ext} B_{ext} = A_{int} B_i \ ,
\end{equation}

\noindent where the areas $A_{ext}$ and $A_{int}$ are the
cross-sectional areas, between the coil and shield and inside the
coil, respectively. From Eqs.(\ref{eq:FieldDifference}) and
(\ref{eq:FluxConservation}) we find:

\begin{equation}
\frac{B_i}{B_{i0}} = \frac{R^2-r_{coil}^2}{R^2} = 1-\left(\frac{r_{coil}}{R}\right)^2 \ ,
\label{eq:HTheoryNormalized}
\end{equation}

\noindent if the width of the coil is negligible.
Eq.(\ref{eq:HTheoryNormalized}) has been normalized to the free
space value, $B_{i0} = \mu_0\lambda$.

\begin{figure}[htb]
        \centering
                \includegraphics[width=8cm]{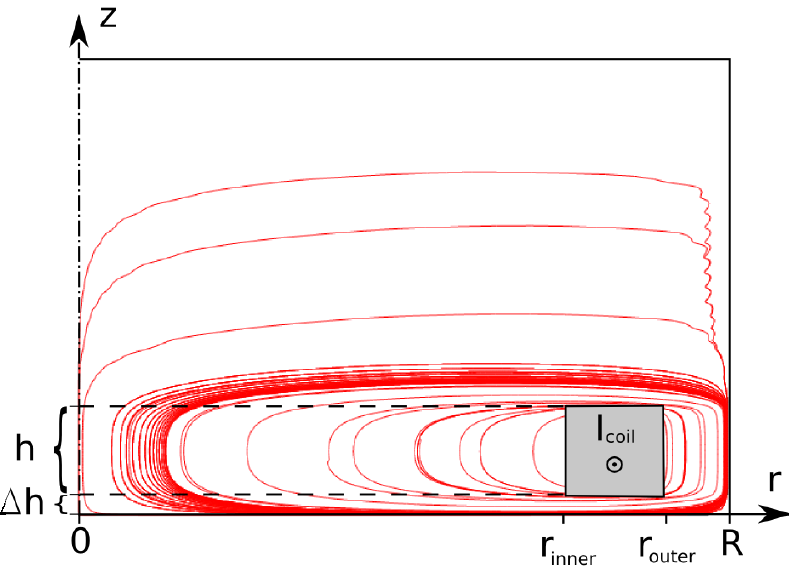}
        \caption{(Color online) Graphical output of a typical simulation. The magnetic field lines are drawn on top of the geometry. Note that in this figure, the z-scale is very compressed compared to the r-scale ($h = 50$mm, $R = 20$mm). Numerical errors introduced by the fieldline algorithm are the cause of the small-lengthscale oscillations of the fieldlines in the top, far right.}
        \label{fig:FieldLines}
\end{figure}

Using Comsol Multiphysics\cite{www.comsol.com} finite-element
magnetostatic simulations, the effect of enclosing a finite solenoid
with a fixed $I_{coil}$ in a superconducting shield has been
investigated. The coil has the parameters $(r_{inner}, r_{outer}, h)
= (15$mm$, 18$mm$, 50$mm$)$, where $h$ is the height (see Fig.
\ref{fig:FieldLines}). The distance between the end of the coil and
the bottom of the superconducting shield can is denoted $\Delta h$.
Furthermore, the current, $I_{coil}$, is DC, which implies a uniform
current density in the coil cross-section.

The problem is axisymmetric and the boundary conditions are set to
magnetic insulation on the boundary of the superconducting shield.
In principle, the shield top should be open, but the simulation is
faster, and the difference in the result is within the error of the
simulation, by setting the top boundary condition to magnetic
insulation as well. The reason is, of course, that it is
sufficiently far away, that only a negligible portion of the
magnetic field lines would go in this area, even if it was open. The
meshing was done automatically, and increasing the mesh density did
not alter results.

The geometry used in the simulation is shown in Fig.
\ref{fig:FieldLines}. The magnetic field lines are drawn on top of
the geometry. The maximum value of the field is located
approximately at the same position, regardless of the coil's
position in relation to the shield.

\begin{figure}[htbf]
        \centering
                \includegraphics[width=8cm]{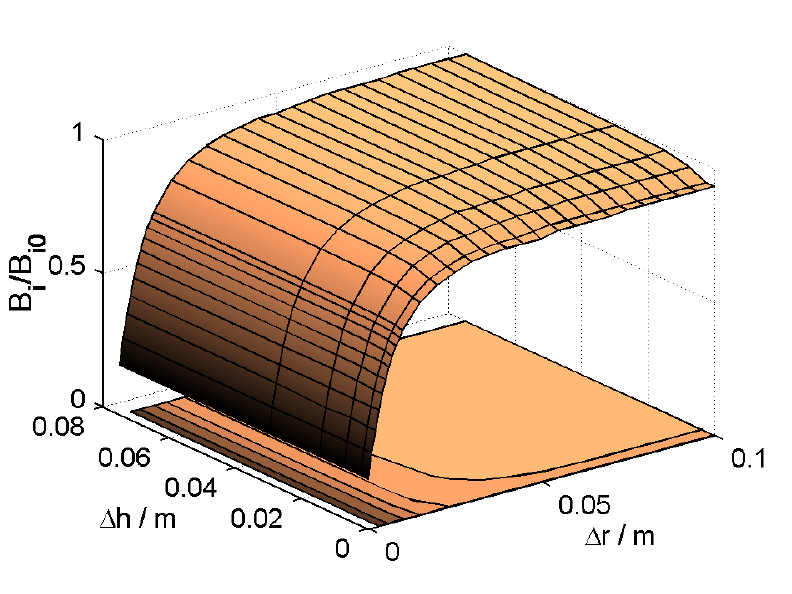}
        \caption{(Color online) Plot of the magnetic field in the center of the solenoid normalized to the free space value as a function ofthe distance to the bottom, $\Delta h$, and radial distance to the superconducting shield, $\Delta r = R - r_{outer}$. The colored plane below the surface is a filled contourplot, which illustrates the shape. The lines in the countourplot represent the numeric solution to the equation $B_{i}(\Delta h, \Delta r) = c$ where $c$ is different for each line.}
        \label{fig:SimulationResults}
\end{figure}

The results of simulations for a large number of geometries are
shown in Fig. \ref{fig:SimulationResults}. The plot shows the
maximum value of the magnetic field strength inside the coil as a
function of the bottom distance, $\Delta h$, and the radial
distance, $\Delta r = R-r_{outer}$, between the solenoid and the
superconducting shield.

The results show a strong influence of the superconducting shield on
the generated magnetic field strength. In fact,
a radial spacing between the solenoid and the superconducting shield
of around $2r_{coil}$ is needed to reach 90\% of  $B_{i0}$.

The deciding factor appears to be the radial distance, $\Delta r$,
as even a rather large $\Delta h$ only gives a 10\% increase in
field. The effect is larger for larger $\Delta r$.

The results show a very weak dependence on $\Delta h$ and thus we
should expect good agreement with the theoretical expression in
Eq.(\ref{eq:HTheoryNormalized}). For fitting purposes, an additional
parameter, $\alpha$, is introduced to deal with the solenoid being
finite, and the nearby capped end of the shield:

\begin{equation}
\frac{B_i}{B_0} = \alpha\frac{R^2-r_{coil}^2}{R^2} \ .
\label{eq:HTheoryNormalizedFit}
\end{equation}

\begin{figure}[htbf]
        \centering
                \includegraphics[width=8cm]{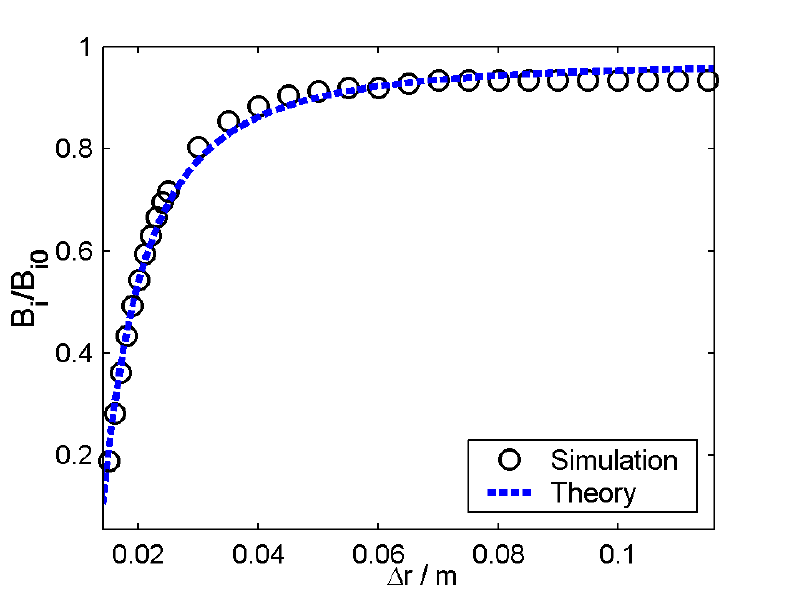}
        \caption{Comparison between a fit of the theoretical expression Eq.(\ref{eq:HTheoryNormalizedFit}) and the simulation output for $\Delta h = 2.5$cm. This fit gives $r_{coil} = 14$mm and $\alpha = 0.94$.}
        \label{fig:fitComparison}
\end{figure}

\noindent Fig. \ref{fig:fitComparison} is a comparison of Eq.
(\ref{eq:HTheoryNormalizedFit}) and the simulation output for one
value of $\Delta h$. The fitting parameters are $r_{coil}$ and the
value of $\alpha$, and the best fit is found for $(r_{coil},\alpha)
= (14$mm$, 0.94)$. Considering the crudeness of the model the fit is
acceptable. It also produces a reasonable value for $r_{coil}$.


The main effect of the cap on the closer end of the shield is to
slightly change the limiting value for $R \rightarrow \infty$, and
thus $\alpha < 1$, as seen in Fig. \ref{fig:SimulationResults}.

The effect of a high-permeability shield inside the superconducting
shield is similar to inserting a large virtual volume of magnetic
vacuum. The virtual volume is a factor of $\mu_r$ thicker than the
actual shielding material, and should thus mediate the effect of
confinement by the superconducting shield. The high-permeability
sheet has been modeled as a cylinder with $\mu_r = 75000$, which is
the stated value for Cryoperm\cite{CryoPermDatasheet}
10$^\registered$ typically used for cryogenic shielding. The result
of inserting a $1$mm thick cylinder between the solenoid and the
superconducting shield is a full recovery of the of the coil factor
to the value obtained for $R \gg r_{coil}$. Also, with the cylinder
inserted, $B_i$ is insensitive to the value of $R$. This is
reasonable, as $1$mm of high-permeability metal with $\mu_r = 75000$
should be roughly equivalent to $75$m of vacuum between the solenoid
and the superconducting shield.

The field strength outside the solenoid can exceed the field
strength inside, when $\Delta r$ is very small compared to
$r_{coil}$. This means, that the critical field of the
superconducting shield might be reached before expected. This may
introduce hysteresis into measurements as well as large trapped
magnetic fields. This can also be countered by the use of a cryoperm
sheet.

In this paper we have presented a commonly overlooked source of
systematic error in cryogenic setups involving magnetic fields. It
was shown, that systematic errors in the coil factor of at least an
order of magnitude can be realized in setups with radial shield
distance $\Delta r \ll r_{coil}$, when comparing to a simple
Hall-probe measurement coil factor or standard free-space formulae.
Furthermore, an approximate theoretical expression was derived for
estimating the real magnetic field or coil factor inside a solenoid
enclosed in a superconducting shield.

The most important parameter is the radial distance between the
solenoid and the shield. The dependence on the distance from the
coil to the shield in the axial direction is very weak - even to the
limit of very small values of $\Delta h$. The solution is to either
make ample space around the coil inside the superconducting shield
or insert a high-permeability metal sheet between the coil and the
superconducting shield. The effect of the sheet is effectively to
insert a virtual vacuum for the magnetic field lines to close in,
thus screening the coil from the effect of confinement. In any case,
this shows the importance of calibrating the solenoid \textit{in
situ}. \textit{In situ} calibration can be done using a SQUID
magnetometer, without a high-permeability metal shield, if the
maximum attainable field in the solenoid is not the limiting factor.

\end{document}